\documentstyle[aps,twocolumn,prl,epsf]{revtex}

\newcommand{\be}{\begin{equation}}
\newcommand{\ee}{\end{equation}}
\newcommand{\bea}{\begin{eqnarray}}
\newcommand{\eea}{\end{eqnarray}}

\begin{document}
\draft

\twocolumn[\hsize\textwidth\columnwidth\hsize\csname @twocolumnfalse\endcsname
\title{Electronic Structure of Lanthanum Hydrides\\
 with Switchable Optical Properties}
\author{K. K. Ng$^{1,2}$, F. C. Zhang$^{1,2}$, V. I. Anisimov$^{2,3}$,
and T. M. Rice$^2$
}
\address{
$^1$Department of Physics, University of Cincinnati,
Cincinnati, Ohio 45221\\ 
$^2$Theoretische Physik, ETH-H\"onggerberg, 8093 Z\"urich, Switzerland\\
$^3$Institute of Metal Physics, Russian Academy of Sciences, 620219,
Ekaterinburg, GSP-170, Russia\\
}
\date{\today}
\maketitle

\widetext
\begin{abstract}
\noindent
Recent dramatic changes in the optical
properties of LaH$_{2+x}$ and YH$_{2+x}$ films discovered by Huiberts
et al. suggest their electronic structure is described best by a local
model. Electron correlation is important in $H^-$-centers and in
explaining the transparent insulating behavior of LaH$_3$. The
metal-insulator transition at $x\sim 0.8$ takes place in a band of
highly localized states centered on the $H$-vacancies in the LaH$_3$
structure. 

\end{abstract}

 ]

\narrowtext
Recently Huiberts et.~al.~\cite{Griessen} reported dramatic changes 
in the optical properties of lanthanum and yttrium hydride films 
with changing hydrogen content, $e.g.$ films that switch 
from a shiny mirror to a yellow transparent window in a time of order seconds.
Although many metal-insulator transitions are known, it is very unusual 
and potentially of technological importance, that the transition leads
to spectacular effects in the visible. Huiberts
et.~al.~\cite{Griessen} point out that    
the electronic structure that underlies this behavior is poorly understood
-- indeed standard LDA-calculations do not predict a metal-insulator
transition  at all~\cite{Dekker,Chou}. In this letter we examine the
role of electron correlation in
these hydrides and argue it justifies a local description that derives
from an  ionic starting point.  

The structural changes are especially small in the $LaH_{2+x}$ films. The
$La$-atoms always form a fcc lattice with  two $H$-atoms always
occupying tetrahedrally coordinated sites. As $x$ changes from $0$ to
$1$, the  octahedrally coordinated site goes from empty to fully
occupied and a good metal ($LaH_2$) evolves to a transparent
insulator ($LaH_3 $). 
The metal-insulator transition in the $d.c.$
conductivity~\cite{Griessen}
occurs at an intermediate concentration, $x \sim 0.8$, far removed
from a lightly doped semiconductor.
This in turn suggests highly localized states associated with $H$-vacancies
in $LaH_3$.
The key challenge is to understand why the structural
changes with changing hydrogen content are remarkably small, but the changes 
in spectral properties are dramatic at energy scales up to a visible
and secondly the origin of the highly localized states at intermediate
concentrations. We shall address these issues in turn. 

We have performed self-consistent LDA calculations on $LaH_2$ and
$LaH_3$. In agreement with previous results~\cite{Dekker} there are
two sets of states -- low energy primarily $1s$-$H$-states and higher
energy states of mainly $5d$-$La$ character. These results suggest
assigning a formal valence $H^-$. In $LaH_2$ the $H^-$-bands are
filled leaving $1 el/f.u.$ in the $5d$-$La$ conduction bands leading to
metallic behaviour~\cite{Dekker}. In $LaH_3$ the $H^-$-bands can hold
all $6 el/f.u.$ but in the LDA calculations there is overlap between the
$H^-$ and $5d$-$La$ bands leading to a semimetal rather than the observed
transparent insulator. 

The LDA calculations give an interesting insight why atomic $H$ is
easily incorporated in this structure.  The net charge change between
$LaH_2$ and $LaH_3$ is found to be remarkably small. The charge at the
octahedral site (within a sphere of 2.533 atomic unit (a.u.) radius) 
in $LaH_3$ is
1.588 $el.$, while the charge in an empty sphere of the same radius in
$LaH_2$ is 0.568 $el.$, so to a very good accuracy, $LaH_{2+x}$ can be
interpreted as neutral $H$-atoms moving into or out of the octahedral
interstitial sites, with negligible changes in the lattice
structure. This charge distribution of $LaH_3$ is not very different
in an ionic picture due to a large radius ($\sim 2 a.u.$) of the outer  
electron in the bound state of the $H^-$-ion.

The $H^-$-ion is a difficult case for LDA and a careful treatment of
the correlation between the two electrons is required~\cite{Bethe,Chandrasekhar,Hylleraas} to obtain the
binding energy $(\simeq$ 0.7 eV). This led us to examine the effect of
correlations on the width of the $H^-$-bands. We start by considering
the hopping integral $t(d)$ for an electron between $H$-atoms with
separation $d$ (in a.u.), ignoring for the moment the $La$-ions. This
can be obtained at once from the lowest energy levels of the
$H_2^-$-ion with inter-proton distance, $d$. $H_2^-$ is a bound
negative ion for $d>3$ -- the range of interest here. 

Let $\psi_i(1,2)$ be the 
groundstate wavefunction  of two electrons in $H^-$, and
 $\phi_i(1)$ be the groundstate
of hydrogen atom.  The three electron states of the $H_2^-$ with
odd(-) and even(+) parities 
 may be constructed 
using the single site (i and j) states:
$\Psi_{\pm} = \Phi_{i,j} \pm \Phi_{j,i},$
where $\Phi_{i,j}=  A [\psi_i(1,2) \phi_j(3)],$
and  $ A$ is an  antisymmetrizer.  The corresponding energies are
given by
$ E_{\pm} = \langle \Psi_{\pm} | h | \Psi_{\pm} \rangle /
\langle \Psi_{\pm} | \Psi_{\pm} \rangle, $
where $ h$ is the Hamiltonian for the $H_2^-$-system, including the 
kinetic  and Coulomb energies~\cite{Fischer}. We choose a variational state of
 Chandrasekhar~\cite{Chandrasekhar} for $H^-$:
\[
\psi(1,2)= (e^{-a r_1-br_2}+e^{-ar_2-br_1})(1+c|\vec r_1-\vec
r_2|)\chi,
\]
where $r$ are the electron radial coordinates, 
$\chi$ is the spinor for a singlet, and  
$a$ = 1.075, $b$ = 0.478, $c$ = 0.312  in a.u..  
This simple wavefunction describes two intraatomic correlated 
electrons well, and its energy is 
very close to the best 
estimate of Hylleraas~\cite{Hylleraas} 
involving 24  variational parameters. The states for
$H_2^-$ thus constructed become exact in the limit $d\gg 1$.
The effective hopping integral  for an electron between 
two $H$-sites can be obtained,
$t(d) = ( E_{-} - E_{+})/2.$
At $d=4$, the groundstate energy estimated in this method is
$ E_{-} = -1.032$ a.u., very close 
to the best available  result of -1.034 a.u. 
 for $H_2^-$  using more complicated 
 variational method~\cite{Taylor}.
 This suggests that the groundstate of 
$H_2^-$  at these separations is very well described by 
a linear combination of $H^-$ and $H$-states, and  justifies our 
approach. 

The two shortest inter-hydrogen distances in $LaH_3$ are those between
$n.n.$ tetra- and octa-sites (denoted as $H_{\rm tet}$ and $H_{\rm
  oct}$) and $n.n.$ tetra-sites which denote as $t_2$ and $t_1$
respectively. These distances are $\sqrt{3}\, a_0/4$ and $a_0/2$
($a_0$: the lattice constant)
respectively (4.58 and 5.29 $a.u.$) giving values $t_2$ = - 0.74 eV
and $t_1$ = -0.54 eV. The $H^-$-bands in tight binding representation
are the eigenstates of the Hamiltonian 
\[
{\cal  H}_{\rm tb} = \sum_{\vec k}{c^{\dagger}(\vec k) M(\vec k) c(\vec
  k)},
\]
where $ c^{\dagger}(\vec k)$ is a 3-component vector, representing
the 3 $H$-states in a unit cell,  and
\[
M(\vec k) =\left(\matrix{\varepsilon_t&t_1\alpha(\vec k)&t_2\beta(\vec k)\cr
             t_1\alpha(\vec k)&\varepsilon_t&t_2\beta^*(\vec k)\cr
        t_2\beta^*(\vec k)&t_2\beta(\vec k)&\varepsilon_o\cr}\right).
\]  
with
\[
\alpha(\vec k)\ =\sum_{\tau =x, y, z}2cos(k_{\tau}/2),
\] 
\[
\beta(\vec k)\ =\sum_{\sigma_1, \sigma_2 =\pm 1}{e^{i(\sigma_1 k_x
+\sigma_2 k_y +\sigma_1 \sigma_2 k_z)/4}}.
\]
Note we now use the lattice parameter $a_0 (5.60 \AA)$
%\stackrel{\;\circ}{A}$) 
as the length unit. The atomic energy levels
at $H_{\rm tet}$ and $H_{\rm oct}$ ($\varepsilon_t$ and
$\varepsilon_0$) we estimate using the LDA calculations as follows. At
the high symmetry
$\Gamma$ point, one can identify in the LDA 
 a 4 by 4 matrix corresponding to the  three 
$H$ and one $La$-6s states
in a unit cell of the fcc lattice.  We can map this matrix onto 
a 3 by 3 submatrix describing the effective three $H$-bands with  the 
same three lower eigenenergies. 
The highest energy band,  representing the $La$-6s state,   is pushed 
well above the bottom of the $La-5d$ conduction band 
due to the strong hybridization, and can be projected out. 
Incorporating the binding state energy of a single $H^-$-ion, which is 
0.7 eV, we obtain $\varepsilon_t=-3.9$ eV, and
$\varepsilon_0=-3.3 $ eV. 

In Fig.~1a the resulting energy bands are plotted. We find a small gap,
which is not sufficient for a transparent
insulator. However we have neglected the $La^{3+}$-ions in
estimating $(t_1, t_2$). In an ionic picture the $La^{3+}$-ions
generate a crystal field at the $H$-sites whose effect can be
estimated as follows. 
Let the interaction between an electron in $H_2^-$ and  the other ions 
be $h'= -\sum_{\vec R} Z(\vec R)e^2/|\vec r -\vec R|$, with $Z=3$ for 
 $La^{3+}$, and Z=-1 for  $H^-$-ions, and 
$\vec R$ the ion position. The energies for $H_2^-$ of two
$H_{tet}$ are modified, due to the crystal field,  to $ E'_{\pm}$, which is
obtained by adding $h'$ to $h$.  The reduction of $t_1$ due to the crystal
field is found
to be $\delta t_1$ = 0.20 eV, where the contribution from within a
cubic unit cell is 0.08 eV.  The estimate for $t_2$ is more
complicated because 
of the  Madelung constants.  Assuming the same percentage reduction,
we estimate the hoppings  
in the presence of the crystal field, $\tilde{t}_1$ = -0.34 eV,
$\tilde{t}_2$ = -0.47 eV.  
The  estimate is based on an ideal ionic approach, and 
neglects the screening effect, so that the actual reduction of  $t$'s
are surely smaller.
With the new values $(\tilde{t}_1, \tilde{t}_2)$, we obtain strongly
narrowed $H^-$-bands (see Fig.~1) and a transparent insulating ground
state. This idealized ionic approach overestimates the narrowing and
neglects the effects of hybridization on the hopping integrals. These
however are weak at the top of the $H^-$-bands and are mainly
important only at energies well below the Fermi energy. 

Turning to the metal-insulator transition in $LaH_{2+x}$ as $x$
increases, we can view it starting either at the metallic $(x=0)$ or
insulating $(x=1)$ endpoints. In the former case, the introduction of
a neutral $H$-atom into a $H_{\rm oct}$-site creates a $s$=1/2 
magnetic impurity, which couples to the conduction $5d-La$ electron spins.
The effective Hamiltonian is a Kondo model,
\[
{\cal H}_{\rm eff}\; =\; \sum_{\vec k, \sigma} {\varepsilon (\vec k) 
d^{\dagger}_{\vec k,\sigma}d_{\vec k, \sigma}} + J \sum_{i} {\vec S_i
\vec s_i}, 
\]
where $i$ runs over all the occupied  $H_{\rm oct}$, and $\vec S$ and
$\vec s$ are the electron spins of the neutral $H$ and of the
conduction electron states. The latter represent linear 
superposition of the $La-5d-e_g$ states around three degenerate
$X_1$ points, which is of $s-wave$ symmetry and can couple to the 
$1s-H$ state.  The antiferromagnetic coupling $J \sim$ 0.7 eV in
free space, and may be renormalized by a numerical factor 
in hydrides. In this Kondo description
conduction electrons are captured by the neutral $H$-atoms at $H_{\rm
  oct}$-sites to form tightly bound singlets and $LaH_3$ is 
viewed as a Kondo insulator with a large band gap.

Starting from insulating $LaH_3$, removing neutral $H$-atoms causes
vacancies at octa-sites $(H_{\rm oct}^V)$ which donate an electron
to the conduction band.
In a conventional semiconductor such as in the doped $Si$, the impurity
state is described by an effective mass theory, and  the result 
is a hydrogen-like bound state with a large effective 
Bohr radius $a_B^*$ of order of hundred $\AA$ due to the light mass and the
large dielectric constant. The critical impurity concentration $x_c$
at which the system becomes metallic is given by Mott criterion 
$x_c \sim (l/a_B^*)^3, $
where $l$ is the interatomic distance. Because of $a_B^* \gg l$, $x_c$
is very small (0.1 percent for $Si$).
The vacancy state in $LaH_{2+x}$ is 
very different, however. Experimentally, it is found that the
semiconducting states extend to $x=0.25$ for lanthanum hydrides, and
to even larger value for 
yttrium hydrides~\cite{Griessen,Shinar}. Below we shall show that 
the vacancy state in lanthanum  hydrides is an
octahedral s-like $La-e_g$ state with extremely small size.  

The $La$-$5d$-$e_g$ and $H_{oct}$ electrons hybridize strongly to form
bonding (mainly $H$) and anitbonding (mainly $e_g$) states. 
A vacancy of $H_{oct}$ breaks the bonds locally and the $e_g$ electron
becomes locally non-bonding, which has much lower energy than the antibonding
$e_g$ states away from the vacancy.  Therefore, in addition to serving as
a positive charge center as in conventional n-type 
semiconductors,  $H_{\rm oct}^V$ creates a potential well for the $e_g$  
state electron. 
The latter is non-perturbative, and is responsible to the 
unusual concentration-dependence of the semiconductor.
   
Consider a $s-$ symmetric octahedral  $e_g$ state ($S$-state hereafter) 
around a  $H_{\rm oct}^V$  as shown in Fig.~2. The $H_{\rm oct}^V$ vacancy
 breaks the bonds on the octahedron and reduces the anitbonding energy
 of the $S$-state. This $S$-state is very localized because
 the neighboring octahedral $s-$ state is antibonding and has a much
 higher energy. We can estimate the depth of the
 potential well, $V_0$, within a perturbation theory upto the second order
in  $V_{sd\sigma}$, 
the hopping integral between an atomic  $d_{3z^2-r^2}$ 
and its neighboring $H_{\rm oct}$ with the orbits towards each other. 
Let $\Delta$ be the atomic
energy difference between $La$-$5$d and $H_{\rm oct}$,
then $V_0=- 6V_{sd\sigma}^2/\Delta $.  This is 2/3  of the
antibonding energy
 $9V_{sd\sigma}^2/\Delta$ at the  $X_1$ point in $LaH_3$.  The latter is estimated 
to be $\sim$ 6 eV from LDA~\cite{Gupta}, so that $V_0 \sim$ -4 eV. 
We may also compare this $S$
state to the non-bonding $t_{2g}$ state. The latter is the lowest energy
conduction state of $LaH_3$, and would be a starting 
point for an effective mass theory in conventional semiconductors.
The $S$ state has much better
kinetic energy as evidenced from LDA for $LaH_2$~\cite{Dekker,Gupta,Harmon}.
A more detailed calculation based on
a tight-binding model of n.n. hopping between $La$ sites estimates 
the kinetic energy gain of the $S$ state over the $t_{2g}$ state is $\sim$
0.9 eV.  This is about to compensate the loss of the antibonding energy
of the $S$ state with the neighboring occupied $H_{\rm oct}$ atoms, which is
$1.5 V_{sd\sigma}^2/\Delta$ $\sim$ 1 eV. 
The $S$ state has  better Coulomb attraction to the
vacancy, because its orbit is towards the central effective charge. 
Therefore the $S$ state should have lower energy than the $t_{2g}$ state, and
is the lowest energy state for 
$H_{\rm oct}$ vacancy. We emphasize that the  potential well generated by the 
$H_{\rm oct}$ vacancy is short-ranged, different from the long-range Coulomb 
force. The effective Bohr radius of the impurity state  is  only half of  
the lattice constant. 

The localized nature of the impurity state in $LaH_{2+x}$ is consistent with 
the temperature dependence of the resistivity data. 
Shinar et.~al.~\cite{Shinar}
reported a phase transition at 
$T$ = 250K, above which the resistivity
shows variable range hopping, below that temperature, it is coherent 
state of the defect band. Localized states are a prerequisite 
for variable range hopping. The coherent state accompanies the onset
of  superlattice order in the $H_{oct}^V$ vacancies. 

The precise nature of the transition in optical properties is not well
understood yet. Since the $d.c.$ conductivity is insulating only 
in the disordered state of $H_{\rm oct}^V$, 
and is metallic in the ordered one, we argue that the transition 
should be a disorder driven Anderson transition, instead of a Mott type
correlation driven transition.

In conclusion, we have studied the electronic structure of $LaH_{2+x}$
starting from 
$H^-$-ionic picture emphasizing the electron
correlation.   Our calculations give an insulator  
for $LaH_3$, in agrement with the recent experiments.
We propose that a $H$-vacancy in $LaH_3$ introduces a $s$-wave  
state centred at the vacant octa-site,with a highly localized form.
This state of the vacancy 
explains why the semiconductor $LaH_{2+x}$ to be so stable in a wide range of 
concentration $x$ ($0.8<x<1$), and is consistent with the experimental
observed variable range hopping mechanism for the resisitivity in
these materials. 
The recent experimental activity of Huiberts et.~al.~\cite{Griessen} has 
raised many interesting questions in  study of metal hydrides.
The precise nature and the criterion of the optical transition from
transparent to reflecting needs be examined. The role of quantum diffusion
of the hydrogen at low temperatures, the order-disorder transitions,
and the interactions between octahedral $H$-vacancies, 
are all interesting open questions.

\begin{figure}
\epsfxsize=3.5in
\epsffile{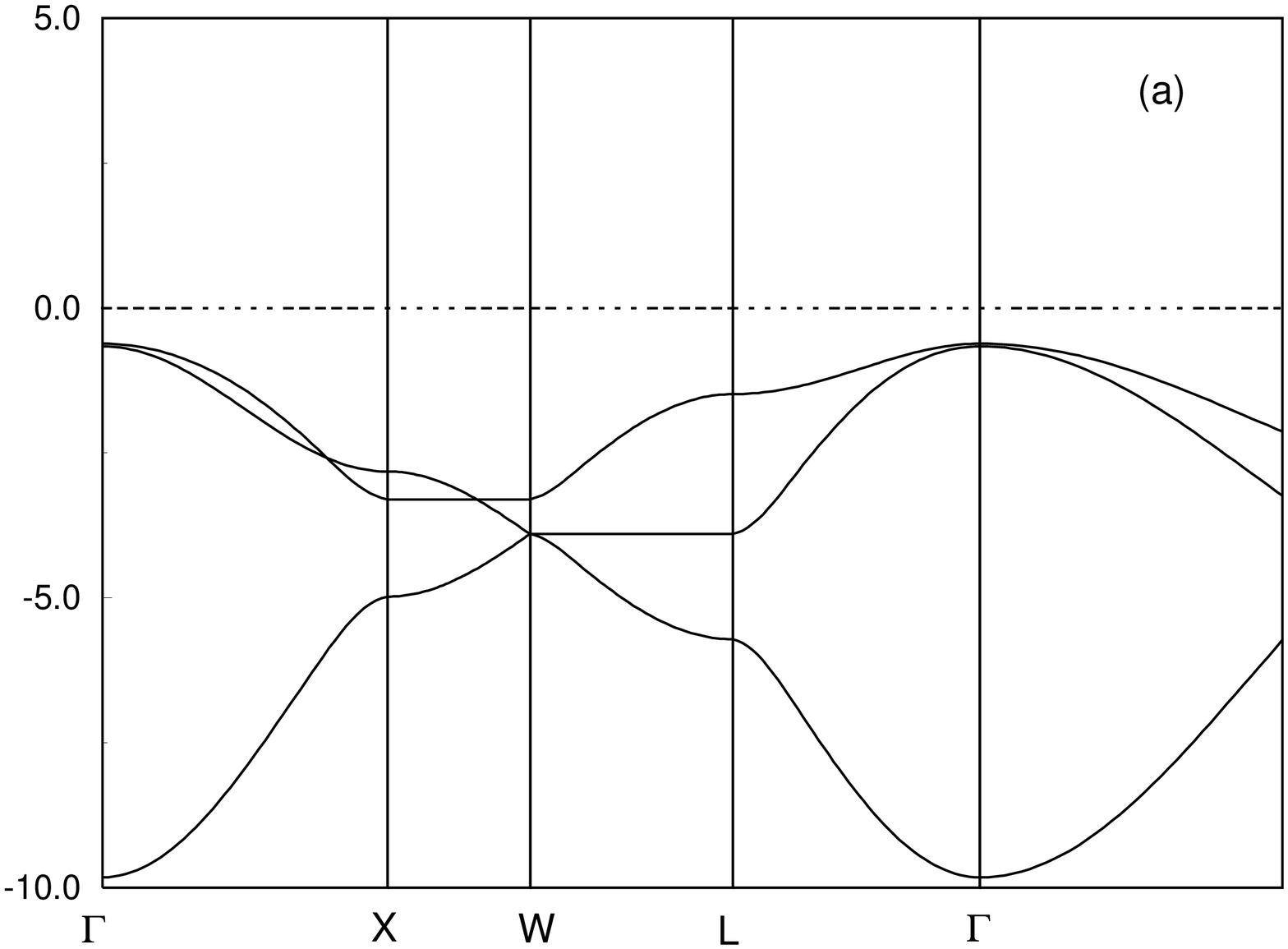}
\epsfxsize=3.5in
\epsffile{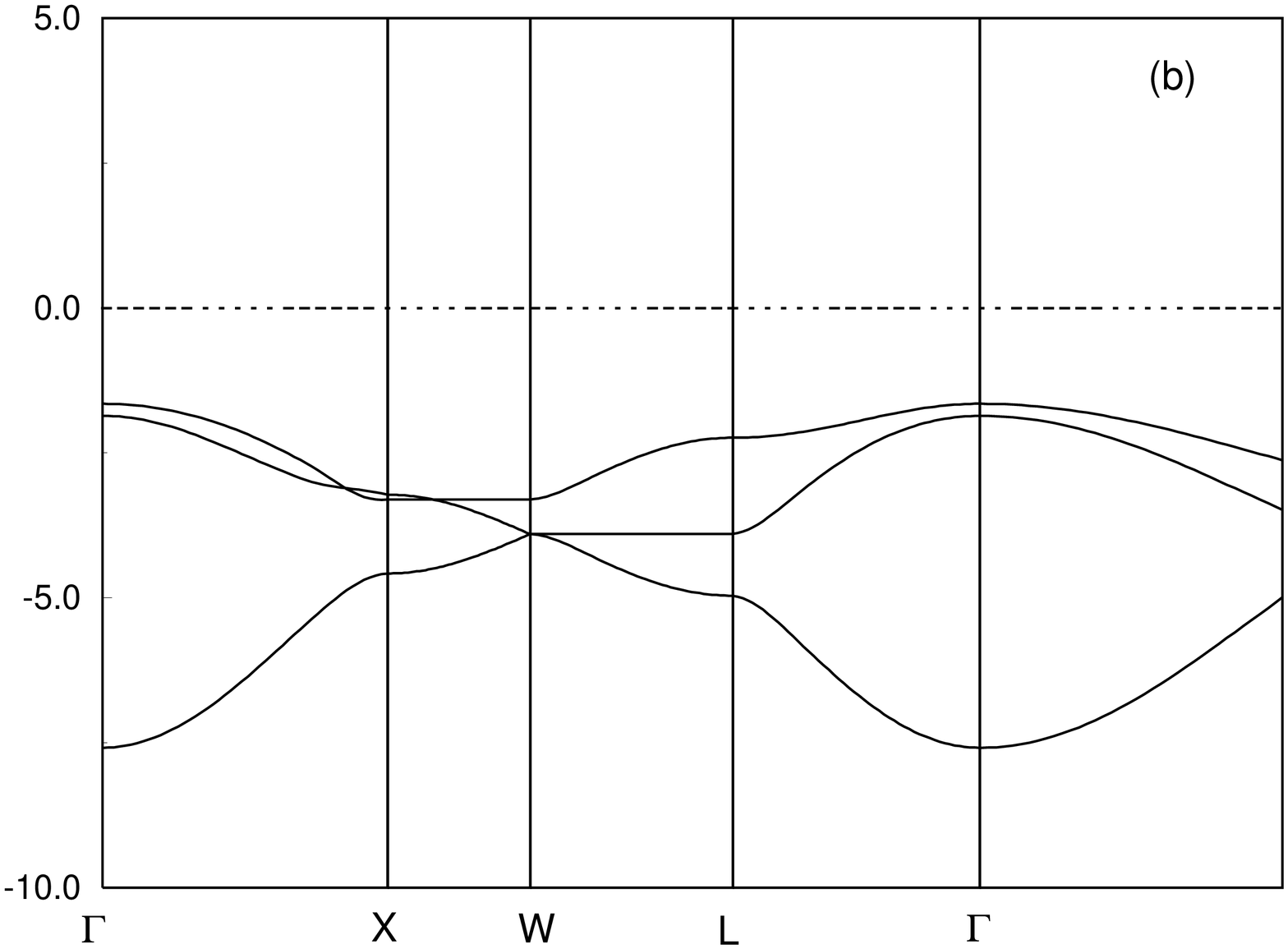}
\caption{The valence bands of $LaH_3$ calculated from the single hole 
hopping in $H^-$. The dashed lines indicate the bottom of the conduction bands.
a): without crystal fields; b): with the crystal fields}
\end{figure}

\begin{figure}
\epsfxsize=3.5in
\epsffile{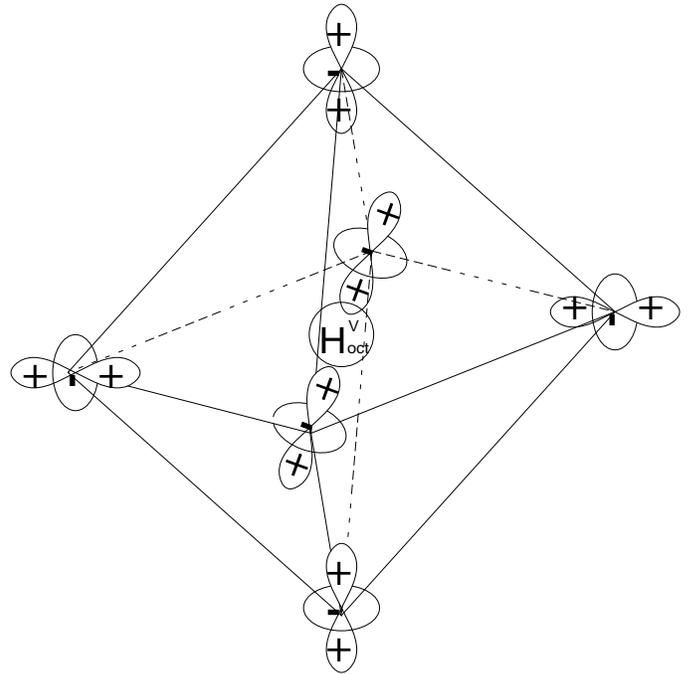}
\caption{Diagrammatic illustration of the proposed  impurity state in
$LaH_3$. The centre circle represents
a missing $H$, forming an  n-type impurity center. The surrounding 
orbits represent phases for the local s-like octahedral La-5d-$e_g$
state.}
\end{figure}

\end{document}